\documentstyle[twoside,psfig,12pt]{article}
\def\double12 {\smallskipamount=6pt plus2pt minus2pt
                  \medskipamount=12pt plus4pt minus4pt
                  \bigskipamount=24pt plus8pt minus8pt
                  \normalbaselineskip=24pt plus0pt minus0pt
                  \normallineskip=2pt
                  \normallineskiplimit=0pt
                  \jot=6pt
                  {\def\smallskip {\vskip\smallskipamount}}
                  {\def\medskip   {\vskip\medskipamount}}
                  {\def\bigskip   {\vskip\bigskipamount}}
                  {\setbox\strutbox=\hbox{\vrule
                    height17.0pt depth7.0pt width 0pt}}
                  \parskip 0pt
                  \normalbaselines}

\def\half12 {\smallskipamount=6pt plus2pt minus2pt
                  \medskipamount=12pt plus4pt minus4pt
                  \bigskipamount=24pt plus8pt minus8pt
                  \normalbaselineskip=16pt plus0pt minus0pt
                  \normallineskip=2pt
                  \normallineskiplimit=0pt
                  \jot=6pt
                  {\def\smallskip {\vskip\smallskipamount}}
                  {\def\medskip   {\vskip\medskipamount}}
                  {\def\bigskip   {\vskip\bigskipamount}}
                  {\setbox\strutbox=\hbox{\vrule
                    height17.0pt depth7.0pt width 0pt}}
                  \parskip 0pt
                  \normalbaselines}

\def\pprint12 {\smallskipamount=4pt plus1pt minus1pt
                  \medskipamount=9pt plus2pt minus2pt
                  \bigskipamount=16pt plus4pt minus4pt
                  \normalbaselineskip=14pt plus0pt minus0pt
                  \normallineskip=1pt
                  \normallineskiplimit=0pt
                  \jot=4pt
                  {\def\smallskip {\vskip\smallskipamount}}
                  {\def\medskip   {\vskip\medskipamount}}
                  {\def\bigskip   {\vskip\bigskipamount}}
                  {\setbox\strutbox=\hbox{\vrule
                   height9.5pt depth4.5pt width 0pt}}
                  \parskip 0pt
                  \normalbaselines}

\def\single12 {\smallskipamount=3pt plus2pt minus2pt
                  \medskipamount=6pt plus4pt minus4pt
                  \bigskipamount=12pt plus8pt minus8pt
                  \normalbaselineskip=12pt plus0pt minus0pt
                  \normallineskip=1pt
                  \normallineskiplimit=0pt
                  \jot=3pt
                  {\def\smallskip {\vskip\smallskipamount}}
                  {\def\medskip   {\vskip\medskipamount}}
                  {\def\bigskip   {\vskip\bigskipamount}}
                  {\setbox\strutbox=\hbox{\vrule
                    height8.5pt depth3.5pt width 0pt}}
                  \parskip 0pt
                  \normalbaselines}

\def\wisk#1{\ifmmode{#1}\else{$#1$}\fi}

\def\le     {\wisk{_<\atop^=}}

\begin{document}
\pagestyle{plain}
\pprint12

\large
\begin{center}
Limits to Global Rotation and Shear \\
From the COBE DMR 4-Year Sky Maps
\end{center}

\medskip
\normalsize
\pprint12
\noindent
\begin{center}
A.~Kogut
\footnote{Electronic Address:
{\tt kogut@stars.gsfc.nasa.gov}} \\
{\it Hughes STX Corporation \\
Laboratory for Astronomy and Solar Physics \\
Code 685, NASA/GSFC, Greenbelt, MD 20771}

\medskip
G. Hinshaw \\
{\it Laboratory for Astronomy and Solar Physics \\
Code 685, NASA/GSFC, Greenbelt, MD 20771}

\medskip
A.J. Banday \\
{\it Max Planck Institut f\"{u}r Astrophysik \\
85740 Garching Bei M\"{u}nchen, Germany}
\end{center}

\medskip
\normalsize
\pprint12
\begin{center}
Accepted for publication in Physical Review D15 \\
\end{center}


\medskip
\begin{center}
\large
ABSTRACT
\end{center}

\normalsize
\noindent
Small departures from a homogeneous isotropic spacetime
create observable features in the large-scale anisotropy
of the cosmic microwave background.
We cross-correlate the
maps of the cosmic microwave background anisotropy
from the Cosmic Background Explorer (COBE) 
Differential Microwave Radiometers (DMR) 
4-year data set
with template maps from Bianchi VII$_{\rm h}$ cosmological models
to limit global rotation or shear in the early universe.
On the largest scales,
spacetime is well described by the 
Friedmann-Robertson-Walker metric,
with departures from isotropy about each spatial point
limited to shear $\sigma / H_0 < 10^{-9}$
and rotation $\omega/H_0 < 6 \times 10^{-8}$
for $0.1 \le \Omega_0 \le 1$.


\renewcommand{\bottomfraction}{0.7}

\clearpage
\section{INTRODUCTION}
One of the goals of cosmology is to understand 
the large scale structure of the universe.  
Gravitational instability models hold that 
large-scale structure forms as the result of 
gravitational amplification of initially small perturbations 
in the primordial density distribution.  
The initial ``seeds'' may result from 
quantum fluctuations in a scalar field (inflation)
or causal ordering following a phase transition 
with broken symmetry (topological defects),
but in all models the evolution occurs in a ``background'' cosmology
described by the Friedmann-Robertson-Walker (FRW) metric.

The anisotropy of the cosmic microwave background (CMB)
provides an observational test of the the assumption 
that on large scales spacetime asymptotically 
approaches the FRW metric.
The detection of fluctuations in the CMB 
provides support for a nearly homogeneous spacetime,
limiting density perturbations 
in the early universe to the level
$\delta \rho / \rho \approx 10^{-5}$.
The CMB anisotropy also provides 
a direct test of the assumption of isotropy
about each spatial point. 
Small deviations from the FRW metric lead 
to observable signatures in the CMB.
Open or flat models with global
rotation or shear will exhibit a spiral pattern of temperature
anisotropy resulting from the handedness of the geodesics propagating
through an anisotropic spacetime.  Further geodesic focusing 
creates ``hot spots'' in open models (density $\Omega_0 < 1$), 
while closed models exhibit a pure quadrupole pattern.

Several authors have used the CMB to limit
rotation $\omega$ and shear $\sigma$ in the universe
\cite{collins_1973,barrow_1985,smoot_1993,bunn_1996}.
Prior to the detection of CMB anisotropy,
Collins \& Hawking 
and Barrow, Juskiewicz, \& Sonoda 
used upper limits on the CMB quadrupole amplitude
to limit $\omega/H_0 < 10^{-5}$
for flat or moderately open ($\Omega_0 \approx 0.3$) models.
Smoot used the quadrupole detection from the 
first-year {\it COBE} DMR sky maps
to limit $\omega/H_0 < 10^{-6}$.
Recently, Bunn, Ferreira, \& Silk 
fitted the full spiral pattern from models with global rotation
to the 4-year DMR data 
and derived limits 
$\omega/H_0 < 3 \times 10^{-7}$.

The DMR anisotropy data are dominated 
by a power-law spectrum of fluctuations
which are ill-described by the pattern of angular anisotropy
predicted for models with global rotation or shear.
Limits to rotation or shear based on full-sky ``template'' maps
must account for the presence of this power-law component.
Bunn, Ferreira, \& Silk 
\cite{bunn_1996}
use a least-squares fit including only instrument noise, 
and then account for chance alignment of the spiral template map 
with features in the power-law component using Monte Carlo simulations.
In this paper, we present an independent analysis using 
a formalism which expressly includes the 
dominant power-law component in the fitting process,
and derive upper limits 
$\omega/H_0 < 6 \times 10^{-8}$ for $0.1 \le \Omega_0 \le 1$.

\section{GEOMETRY}
The FRW geometry is a special case of 
more general solutions to the Einstein field equations.  
Relaxing the requirement of isotropy about each point 
leads to more complicated solutions which contain 
the FRW metric as a special case.  
The Bianchi models VII$_{\rm h}$, VII$_{\rm 0}$, and IX are the most
general solutions for a homogeneous but anisotropic universe which
contain the open, flat, or closed FRW metric as a special case.   The
metric is defined as
$$
g_{\mu \nu} = - {\rm n}_\mu {\rm n}_nu ~ + ~g_{\rm AB} 
{\rm E}^{\rm A}_\mu {\rm E}^{\rm B}_\nu
$$
where n$_\mu = -t_{;\mu}$ is 
the normal to the spacelike surfaces of homogeneity, 
$g_{\rm AB}$ is a $3 \times 3$ matrix 
depending only on t, 
and E$^{\rm A}_\mu$ are the three invariant covector fields in the 
surfaces of homogeneity such that 
$$
{\rm E}^{\rm A}_{\mu ; \nu} - {\rm E}^{\rm A}_{\nu ; \mu} =
{\rm C}^{\rm A}_{\rm BC} {\rm E}^{\rm B}_\mu {\rm E}^{\rm C}_\nu 
$$
where 
C$^{\rm A}_{\rm BC}$ are the structure constants 
(see, e.g., Ellis \& MacCallum
\cite{ellis_1969}).
The matrix $g_{\rm AB}$ can be expanded into 
a part representing the expansion 
and a trace-free part representing the anisotropy,
$$
g_{\rm AB} = {\rm e}^{2a} [{\rm e}^{2 \beta}]_{\rm AB}.
$$
The metric anisotropy in these models is characterized 
by the vorticity $\omega$ and the shear $\sigma$
(differential expansion in orthogonal directions), 
$$
\omega^\mu = \frac{1}{2} \eta^{\mu nu \lambda \rho} 
{\rm u}_\nu {\rm u}_{\lambda ; \rho}
$$
$$
\sigma_{\rm ij} = \frac{1}{2} 
\left[ (\exp({\beta}))^\bullet (\exp({- \beta})) +
  (\exp({- \beta})) (\exp({\beta}))^\bullet \right]_{\rm ij}
$$
where u is the four-velocity, 
$\eta^{\mu \nu \lambda \rho}$ is an anti symmetric tensor 
and the dot represents a time derivative.  
The FRW metric is recovered in the limit $\omega/H_0 \rightarrow 0$
and $\sigma/H_0 \rightarrow 0$
where $H_0$ is the Hubble constant.  

Photons propagate along geodesics from the surface of last scattering
to the observer; for small metric anisotropies the CMB anisotropy is
$$
\frac{\Delta T}{T} = ({\rm p}^{\rm i}{\rm u}^{\rm i})_{\rm R}
- ({\rm p}^{\rm i}{\rm u}^{\rm i})_{\rm E}
- \int_{\rm R}^{\rm E} {\rm p}^{\rm j}{\rm p}^{\rm k} 
  \sigma_{\rm jk} dt
$$
where p$^{\rm i}$ are the direction cosines of the null geodesic 
and the subscripts E and R refer to 
emission and reception, respectively.  
The first two terms are the Doppler anisotropy 
from the motion of the receiver and the surface of last scattering,
while the last term represents the geodesics integrated over the
anisotropic metric. Various authors have solved the geodesic equations
for the observed pattern of CMB anisotropy
\cite{collins_1973,barrow_1985,novikov_1968}.
In this paper, we specialize to Bianchi VII$_{\rm h}$ model
($C^2_{31} = C^3_{21} = 1$ and $C^2_{21} = C^3_{31} = \sqrt{h}$),
which contains the Bianchi I, V, and VII$_{\rm 0}$ models 
in the appropriate limit.
Two effects are important.  The intrinsic handedness of
the tensor $\sigma_{\rm ij}$ 
causes the geodesics to spiral in open or flat models.
The resulting temperature pattern is given by
\begin{equation}
\frac{\Delta T}{T} ~= ~\left( \frac{\sigma}{H} \right)_0 [
~A(\theta_{\rm R}) \sin \phi_{\rm R} ~+ 
~B(\theta_{\rm R}) \cos \phi_{\rm R} ]
\label{spiral_eq}
\end{equation}
(Barrow, Juskiewicz, and Sonoda \cite{barrow_1985}),
where
($\theta_{\rm R}, \phi_{\rm R}$) specify the 
photon direction\footnote{
The direction an observer sees on the sky is given by 
$\theta = \pi - \theta_{\rm R}, ~\phi = \pi + \phi_{\rm R}$.},
\begin{eqnarray}
A(\theta_{\rm R}) & = &
C_1 [ \sin \theta_{\rm R} 
- C_2( \cos \psi_E + 3 \sqrt{h} \sin \psi_E) ] \nonumber \\
 & & + ~C_3 
\int^{\tau_{\rm R}}_{\tau_E}
\frac{s (1-s^2) \sin \psi d\tau}
{(1+s^2)^2 {\rm sinh}^4(\sqrt{h} ~\tau/2)}, 
\label{a_eq}
\end{eqnarray}
\begin{eqnarray}
B(\theta_{\rm R}) & = &
C_1 [ 3\sqrt{h} \sin \theta_{\rm R} 
- C_2( \sin \psi_E + 3 \sqrt{h} \cos \psi_E) ] \nonumber \\
  & & - ~C_3 
\int^{\tau_{\rm R}}_{\tau_E}
\frac{s (1-s^2) \cos \psi d\tau}
{(1+s^2)^2 {\rm sinh}^4(\sqrt{h} ~\tau/2)}, 
\label{b_eq}
\end{eqnarray}
\begin{equation}
s = \tan \left(\frac{\theta_{\rm R}}{2} \right)
 \exp[ -\sqrt{h}(\tau - \tau_{\rm R}) ],
\label{s_eq}
\end{equation}
\begin{equation}
\psi = (\tau - \tau_{\rm R}) - \frac{1}{\sqrt{h}} \ln \left\{
\sin^2 \left( \frac{\theta_{\rm R}}{2} \right) + 
\exp[2\sqrt{h}(\tau - \tau_{\rm R})] 
\cos^2 \left( \frac{\theta_{\rm R}}{2} \right) \right\},
\end{equation}
and $dt = {\rm e}^a d\tau$. 
The constants $C_1$, $C_2$, and $C_3$ are defined as
\begin{eqnarray}
C_1 = \frac{1}{3 \Omega_0 x} \nonumber \\
C_2 = \frac{2 s_E (1 + z_E)}{1 + s^2_E} \nonumber \\
C_3 = \frac{4 \sqrt{h} (1 - \Omega_0)^{3/2}}{\Omega_0^2}, 
\label{c_eq}
\end{eqnarray}
where $z$ is redshift and the parameter
$$
x = \sqrt{\frac{h}{1 - \Omega_0}}
$$
determines the scale over which the principal axes of shear and
rotation change orientation.
We obtain the Bianchi VII$_{\rm 0}$ model in the limit 
$\Omega \rightarrow 1, h \rightarrow 0$ with $x$ finite;
we obtain the Bianchi I model as $x \rightarrow \infty$
and the Bianchi V model as $h \rightarrow \infty$.

\begin{figure}[b]
\centerline{
\psfig{file=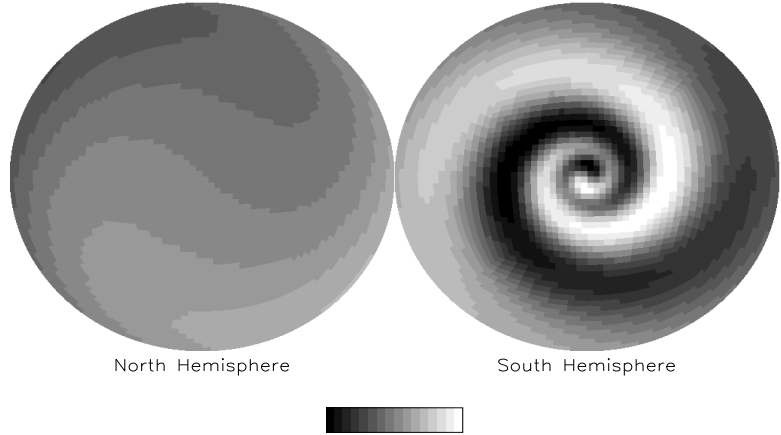,width=4.5in,height=3.0in,angle=90}}
\caption{Anisotropy template map (arbitrary units) 
for a Bianchi VII$_{\rm h}$ model
with $\Omega_0 = 0.3$ and $x = 0.5$.
Note the differences between the hemispheres from geodesic focusing.}
\label{spiral_fig}
\end{figure}

The ratio $\sigma/H_0$ determines the amplitude of the anisotropy,
while the parameter $x$ determines the pitch angle of the spiral.  
A significant amount of the resulting CMB anisotropy is present 
at high-order moments than the quadrupole.  
A second effect is geodesic focusing, present only in open models. 
The component of the geodesics along the symmetry axis transforms 
in these models as
$$
\tan\left( \frac{\theta}{2} \right) = 
\tan\left( \frac{\theta_{\rm R}}{2}
~\exp(\tau_{\rm R} - \tau) \right).
$$
Since
$$
\sinh\left( \frac{\tau}{2} \right) 
~\approx ~\frac{\Omega^{-1} - 1}{1 + z},
$$
the spiral anisotropy discussed above is focused 
into a single hot ``navel'' 
$\approx ~\Omega_0$ radians in scale 
oriented along the symmetry axis of the metric.  
Figure \ref{spiral_fig} shows a polar projection 
of the anisotropy for $\Omega_0 = 0.3$ and $x = 0.5$.

\section{ANALYSIS}
The DMR 4-year sky maps are well described by a superposition of
instrument noise and nearly scale-invariant CMB anisotropy,
in the sense that the observed sky falls near the median of
the distribution of simulated skies 
which include only these two components
\cite{dmr_median}.
We test for the presence of a third component of unknown amplitude
whose spatial distribution is described 
by a Bianchi VII$_{\rm h}$ model,
$$
\Delta T^{\rm DMR} =  \Delta T^{CMB} ~+ ~\alpha \Delta X ~+ ~n,
$$
where $\Delta T^{\rm CMB}$ is the thermodynamic temperature of 
the scale-invariant CMB anisotropy,
$\Delta X$ is a ``template'' map of a specific Bianchi model, 
and  $n$ is the DMR instrument noise.
We do not analyze the closed Bianchi IX model,
which exhibits neither spiral anisotropy nor geodesic focusing.
We use a weighted combination of the 6 channel maps 
from the DMR 4-year data,
from which a model of Galactic emission has been removed
(the ``correlation'' map of Refs. 
\cite{kogut_1996a,hinshaw_1996}).
To further reduce residual Galactic contamination,
we analyze only the high-latitude portion of the resulting map,
here defined as latitude $|b| > 20^\circ$ with custom cutouts
at Ophiuchus and Orion
\cite{galcut_citation}.
For computational efficiency,
we degrade the maps one step in pixel resolution,
leaving 954 high-latitude pixels,
each of size $5{{\rlap.}^\circ}2.$

We estimate the correlation coefficient $\alpha$ 
by minimizing
\begin{equation}
\chi^2 ~= \sum_{a,b} 
~(T - \alpha X)_a
~({\bf M}^{-1})_{ab}
~(T - \alpha X)_b,
\label{chisq_eq}
\end{equation}
where
$T$ is a vector consisting of the DMR temperatures 
in each high-latitude pixel,
$X$ is a similar vector for the template map,
and ${\bf M}$ is the covariance matrix between the elements of $T$.
Several authors have demonstrated that 
$\Delta T^{\rm CMB}$ can adequately be described as 
a Gaussian random field
with a scale-free power spectrum
\cite{dmr_median}.
We assume that the Gaussian component
is uncorrelated with the geodesic component $\Delta X$.
The covariance matrix (including instrument noise) thus becomes 
\begin{equation}
M_{ab} 	= \langle T_a T_b \rangle
	= \frac{1}{4 \pi} \sum_\ell (2\ell + 1) W^2_\ell C_\ell 
	P_\ell(\hat{n_a} \cdot \hat{n_b})
	+ \frac{\sigma_0^2}{N_a} \delta_{ab},
\label{covar_eq}
\end{equation}
and is a function of the instrument noise 
and power spectrum $C_{\ell}$,
where 
$W^{2}_{\ell}$ is the experimental window function 
that includes the effects of beam smoothing and finite pixel size,
$P_l(\hat{n_a} \cdot \hat{n_b})$ 
is the Legendre polynomial of order $\ell$, 
$\hat{n_a}$ is the unit vector towards the center of pixel $a$,
and $N_a$ is the number of observations for pixel $a$.
In the limit that only instrument noise is considered 
($Q_{rms-PS} = 0$),
the covariance matrix ${\bf M}$ is diagonal and
Eq. (\ref{chisq_eq}) reduces to a noise-weighted least-squares 
estimate of $\alpha$.

We adopt a scale-invariant power spectrum
\cite{bond_1987},
\begin{equation}
C_{\ell} = C_{\ell}(Q_{rms-PS},n) \equiv
	(4\pi/5)Q_{rms-PS}^{2} \,  
	\frac{\Gamma(\ell+(n-1)/2)\Gamma((9-n)/2)}
	     {\Gamma(\ell+(5-n)/2)\Gamma((3+n)/2)},
\label{qn_model_eq}
\end{equation}
with amplitude $Q_{rms-PS} = 18 ~\mu$K
~and index $n=1$,
and set $C_0 = C_1 = \infty$
to assign zero weight to the monopole and dipole terms
(in practice, we use $C_0 = C_1 = 10^8 ~\mu{\rm K}^2$).
Strictly speaking,
the power spectrum in Eq. (\ref{qn_model_eq})
represents a flat cosmology,
while the Bianchi VII$_{\rm h}$ template maps
represent open models.
Although our choice of an exact scale-invariant flat model
may not represent the true CMB distribution in detail,
it is certainly adequate for these purposes:
a variety of analyses have shown that the DMR data 
can not discriminate at high confidence 
between a scale-invariant power spectrum
and alternate (e.g., open) models 
\cite{dmr_models}.
Our results are not dependent on the detailed model normalization:
using $Q_{rms-PS} = 24 ~\mu$K and $n=0$
changes our limits by only 20\%.

Equation (\ref{chisq_eq}) 
provides an analytic solution for the coupling coefficient $\alpha$
between the template map $X$ and the DMR data,
\begin{equation}
\alpha = \frac{\sum_{a,b} ~T_a ~({\bf M}^{-1})_{ab} X_b}
              {\sum_{a,b} ~X_a ~({\bf M}^{-1})_{ab} X_b},
\label{alpha_eq}
\end{equation}
assuming a specific orientation of the two maps.
In practice, the relative orientation of the template is unknown.
For a selected Bianchi model
specified by the parameters [$\Omega_0, x$],
we create a template map by numerical integration of
Eqs. (\ref{spiral_eq}) -- (\ref{c_eq}),
and find the fitted value of the correlation coefficient $\alpha$
and its uncertainty $\delta \alpha$
in each of 13824 different orientations
on an approximately $10^\circ$ angular grid: 
we step the symmetry axis 
to point to each of the 384 unit vectors
on a quadrilateralized sky cube
\cite{white_1992}
at resolution index 4,
then rotate the template by $10^\circ$ steps 
in azimuth about each new symmetry axis.
The uncertainty  
$\delta \alpha = 
\sum_{a,b} \left[ X_a ({\bf M}^{-1})_{ab} X_b \right]^{-1/2}$
varies slightly with angular orientation
as different features in the template map $X$
move into and out of the 
low-latitude region excluded by the Galactic cut.
Accordingly, we use the parameter 
$\Gamma = \alpha / \delta \alpha$
to assess the statistical significance 
of the correlation in each specific orientation,
and select the largest $|\Gamma|$ to denote the ``best'' orientation.

\begin{figure}[b]
\centerline{
\psfig{file=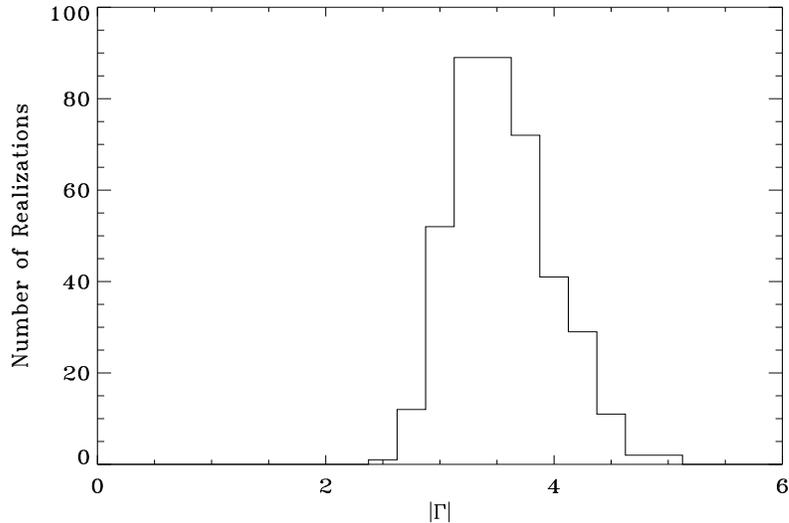,width=4.5in,height=3.0in,angle=90}}
\caption{Histogram of parameter $\Gamma = \alpha / \delta \alpha$
with input $\alpha = 0$.
Chance alignments produce a ``best fitted'' value
between 3 and 5 standard deviations from zero.}
\label{gamma_fig}
\end{figure}

Even in the absence of geodesic effects in the CMB,
chance alignments between the Bianchi template map 
and the power-law component of CMB anisotropy 
can create non-zero correlations.
If the relative orientation of the DMR and template maps
is held constant,
$\Gamma$ will follow a normal distribution 
with zero mean and unit variance
over an ensemble of CMB maps.
By selecting the largest $|\Gamma|$
from a large number of possible orientations 
in a {\it single} CMB map
with a {\it fixed} pattern of noise and anisotropy,
we maximize $|\Gamma|$ over an ensemble of chance alignments
and thus expect the largest $|\Gamma|$
to lie on the tail of a statistical distribution.

We test the accuracy of the fitting process
and assess the statistical significance of the results
using Monte Carlo simulations.
For each choice of Bianchi parameters [$\Omega_0, x$],
we synthesize 100 realizations 
of the power-law component of CMB anisotropy
[Eq. (\ref{qn_model_eq})],
add instrument noise following the DMR 4-year observation pattern,
and derive the best fitted correlation coefficient $\alpha$
from the maximum of $|\Gamma|$ as described above.
The distribution in fitted orientation is flat:
there is no preferred direction for the symmetry axis
as might be caused by the Galactic cut.
Figure \ref{gamma_fig}
shows the distribution in $|\Gamma|$. 
Most of the realizations have $|\Gamma| \approx 3.7$;
95\% of the realizations have $|\Gamma| < 4.5$, 
independent of the parameters [$\Omega_0, x$].

We also synthesize additional realizations
which include a component $\alpha_0 X$ 
with $\Gamma = 7$ to be safely above the noise.
When the input template $X$ is exactly aligned on our angular grid,
the distribution of fitted $\alpha$ has 
mean $\langle \alpha \rangle = \alpha_0$,
and correctly selects the input orientation.
When the input template is misaligned,
we recover a slightly reduced amplitude
$\langle \alpha \rangle = 0.9 \alpha_0$
with orientation correctly centered on the closest grid point.

\section{RESULTS}
We find no statistically significant correlations
between the DMR 4-year CMB map
and any Bianchi VII$_{\rm h}$ model with
$0.1 \le \Omega \le 1$ and $0.1 \le x \le 10$.
The most significant correlation occurs
at $\Omega_0 = 0.15$, $x=0.2$,
for which $\Gamma = 4.37$.
A coefficient this large occurs in 7\% of the null simulations. 
The minimum value occurs 
at $\Omega_0 = 0.95$, $x=2.0$,
for which $\Gamma = 2.91$ (1\% of simulations).
The range of fitted coefficients $\alpha$ 
over the grid of values [$\Omega_0, x$]
agrees well with the distribution expected from chance alignments
of instrument noise and power-law CMB anisotropy:
we find no compelling evidence for the presence of a third
component traced by any Bianchi  VII$_{\rm h}$ model.
We thus adopt the value $\Gamma = 4.5$
at each grid point in [$\Omega_0, x$]
as a 95\% confidence upper limit to the amplitude of 
microwave anisotropy traced by the corresponding Bianchi template.

\begin{figure}[b]
\centerline{
\psfig{file=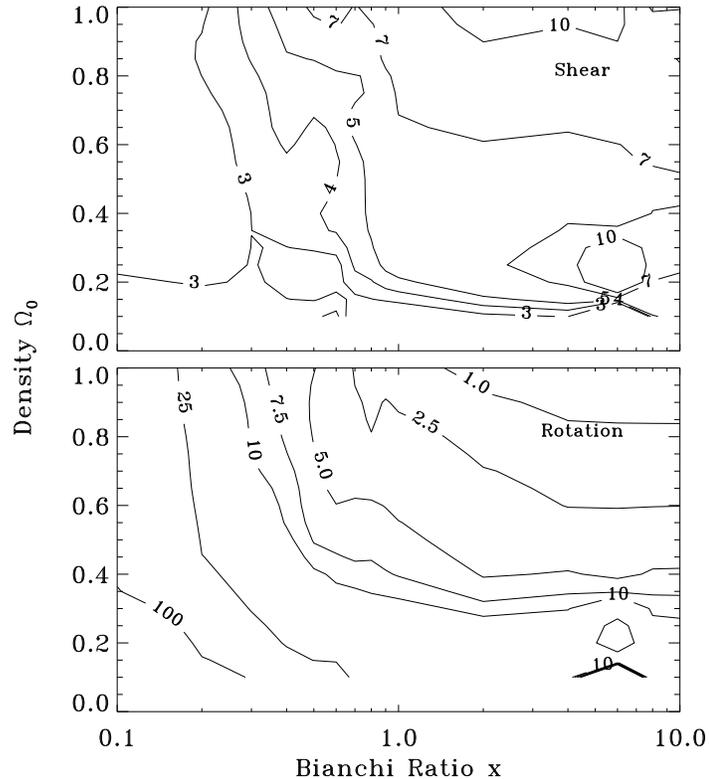,height=4.5in}}
\caption{95\% confidence level upper limits to shear and rotation 
from the DMR 4-year CMB map.
(top) Shear $\sigma / H_0 \times 10^{10}$.
(bottom) Rotation $\omega / H_0 \times 10^{10}$.
 }
\label{limits_fig}
\end{figure}

Figure \ref{limits_fig} shows the resulting limits
to shear $\sigma$ and rotation $\omega$.
In the limit $x \rightarrow \infty$,
the template pattern approaches a pure quadrupole
and $\sigma / H_0$ 
is limited by confusion with the observed quadrupole amplitude
and quadrupolar emission from the Galaxy.
For $x < 2$ the spiral pattern dominates
and we obtain a limit
$\sigma / H_0 < 10^{-9}$,
corresponding to rotation
$\omega/H_0 < 5 \times 10^{-9}$ for flat models and 
$\omega/H_0 < 6 \times 10^{-8}$ for open models.

Our limits on $\sigma / H_0$ are approximately 
a factor of three below those of Bunn, Ferreira, \& Silk. 
This results primarily from our inclusion 
of the power-law component of the CMB anisotropy 
in the covariance matrix ${\bf M}$,
which accounts for the chance alignments expected 
(in an ensemble average sense)
between the template and power-law components.
Bunn, Ferreira, \& Silk use a noise-weighted fit,
\begin{equation}
\alpha = \frac{ \sum_i (T_i X_i) / \sigma_i^2 }
	      { \sum X_i^2 / \sigma_i^2}
\end{equation}
({\it cf} Eq. (7) in Ref. \cite{bunn_1996}),
and set limits on cosmological parameters by 
comparing the value from the DMR data to the distribution 
of values from Monte Carlo simulations with $\alpha=0$.
We may reproduce this technique in our analysis 
by keeping only the noise term in Eq. (\ref{covar_eq})
and examining the resulting distribution in $|\Gamma|$
for Monte Carlo simulations that 
include the power-law contribution in the sky maps.
If the power-law contributions are neglected in the fitting process,
the distribution of $|\Gamma|$ shifts to larger values
and becomes substantially broader,
resulting in weaker limits to the rotation and shear.
If we repeat our results 
using only the noise term in Eq. (\ref{covar_eq}),
we obtain weaker limits compatible with those of 
Bunn, Ferreira, \& Silk. 

We conclude that 
the observed microwave sky 
shows no evidence for geodesic effects
caused by an anisotropic spacetime.
On the largest scales,
spacetime is well described by the FRW metric,
with deviations from homogeneity $\delta \rho /\rho \approx 10^{-5}$
and departures from isotropy $\sigma / H_0 < 10^{-9}$.

\large
\begin{center}
ACKNOWLEDGMENTS
\end{center}

\normalsize
This work was funded by NASA grant S-57778-F.


\end{document}